\documentclass[aps,twocolumn,floatfix,showpacs]{revtex4-1}
\usepackage{amsmath,graphicx,color,mathrsfs,bm}
\begin{document}
\def\be{\begin{equation}}
\def\ee{\end{equation}}
\def\bea{\begin{eqnarray}}
\def\eea{\end{eqnarray}}
\newcommand{\ver}{{\bf r}}
\newcommand{\ttime}{{\cal T}}

\title{Importance-sampling computation of statistical properties of
coupled oscillators}
\author{Shamik Gupta$^1$, Jorge C. Leit\~{a}o$^2$, and Eduardo G.
Altmann$^{2,3}$}
\affiliation{$^1$Department of
Physics, Ramakrishna Mission Vivekananda University, Belur Math, 711 202
Howrah, India \\ $^2$Max Planck Institute for the Physics of Complex
Systems, D-01187 Dresden, Germany \\ $^3$School of Mathematics and Statistics, University of Sydney, 2006 NSW, Australia}
\begin{abstract}
We introduce and implement an importance-sampling Monte Carlo algorithm to study
systems of globally-coupled oscillators. Our computational method efficiently obtains estimates of the tails of the
distribution of various measures of dynamical trajectories corresponding
to states occurring with (exponentially) small probabilities. We demonstrate the general validity of our results by applying the method to two 
contrasting cases: the driven-dissipative Kuramoto model, a paradigm in the study of spontaneous
synchronization; and the conservative Hamiltonian mean-field model, a
prototypical system of
long-range interactions. We present results for the distribution of 
the finite-time Lyapunov exponent and a time-averaged order parameter.
Among other features, our results show most notably that the
distributions exhibit a vanishing standard deviation but a skewness that
is increasing in magnitude with the number of oscillators,
implying that non-trivial asymmetries and states yielding rare/atypical values of
the observables persist even for a large number of
oscillators.
\end{abstract}
\date{\today}
\maketitle
\section{Introduction}
\label{sec:intro}

Non-linear dynamical systems often exhibit different behavior in different regions of the phase space.
A prominent example is the Fermi-Pasta-Ulam chain of non-linear
oscillators, for which a variation in the initial
condition leads to physically very different long-time solutions, e.g.,
solitons and chaotic breathers \cite{Gallavotti:2007}. Another example is the generic co-occurrence of regular (quasiperiodic) and chaotic trajectories in Hamiltonian systems~\cite{MacKay:1987}.
Even when trajectories leading to one type of behavior are 
rare or atypical, they may still be responsible for important physical
phenomena. An example is that of chemical reactions in which the system
goes over from being constituted with one chemical species to another, due to atypical trajectories passing through unstable saddle point regions
in the phase space \cite{Komatsuzaki:2002}. It is thus of great interest to identify and
characterize the variation in dynamical evolution with respect to initial conditions
and the occurrence of atypical dynamical trajectories in many-body
non-linear systems.

A different motivation to study atypical trajectories and dependence on
initial conditions arises from a theorist's
endeavor to reconcile analytical calculations done in the thermodynamic
limit, $N \to \infty$, with numerical simulations for a finite number $N$
of particles. Calculations often assume that for generic dynamical
systems, different initial conditions have similar behavior, which may be justified on
heuristic grounds in the limit $N \to \infty$. A pertinent question for
finite $N$ is however to ask to what extent the infinite-$N$ scenario holds and what is the role
of initial conditions. Specifically, one may ask for a
finite-$N$ system: Do most initial conditions behave similarly to that
predicted in theory, so that the latter is indeed the typical behavior
for a finite system? Or is there a significant fraction of initial
conditions that show large deviations with respect to the theoretically
predicted behavior? What is the relative fraction of these two types of
behavior (atypical vs. typical), and how does this fraction approach
zero (if at all) as $N\rightarrow \infty$?
While a resolution of these issues helps to view objectively analytical predictions {\em vis-\`{a}-vis}
numerical simulations, one obtains as an offshoot a quantitative view of the variability of phase
space regions.

In this work, we address the aforementioned issues in a system of coupled oscillators, which offers a useful framework to connect low-dimensional systems studied in the field of non-linear dynamics with the high-dimensional ones dealt with in statistical
mechanics \cite{Strogatz:2014}. In specific limits, the system we
consider describes the equilibrium, second-order, conservative dynamics of a
paradigmatic long-range interacting system, and a non-equilibrium,
first-order, dissipative dynamics in the presence of disorder that describes a prototypical model of
spontaneous collective synchronization. Two specific observables that we focus
on to bring out the dependence on initial conditions for finite
systems are (i) the finite-time Lyapunov exponent (FTLE), and (ii) a
time-averaged order parameter, both
quantifying the 
variability in the finite-time dynamical evolution of trajectories
starting from different initial conditions. 

The computational challenge we tackle in this paper is the efficient estimation 
of the probability of an observable computed over an ensemble of states,
and in particular, the probability of occurrence of atypical values away from the typical
value of the observable. Estimation of probability of rare events is a
traditional problem in statistical physics (tackled, e.g., using Monte
Carlo methods~\cite{Newman:2002}). In the last decade, the issue of rare-event simulation has seen a
surge in interest in a number of varied contexts involving non-linear
dynamical systems \cite{Kurchan:2007}. 
In this work, we accomplish the desired task of computation of
probability by starting from the framework of a Metropolis-Hastings
 Monte Carlo method proposed in Ref.~\cite{Leitao:2017} for rare-event
 simulation in simple chaotic maps, and by suitably adapting the
 method to apply to the more complex dynamics of coupled oscillators. Our method allows us to study systems involving thousands
 of coupled oscillators, in contrast to small systems ($\sim 32$ degrees
 of freedom) considered previously~\cite{Kurchan:2007,Leitao:2017}.
A successful implementation of our algorithm demonstrates the generality
of our method and results with respect to the two different observables,
namely, the FTLE and the time-averaged order parameter, and with respect
to two very different and contrasting dynamics, namely, an equilibrium,
second-order, conservative dynamics, and a non-equilibrium, first-order,
dissipative dynamics in presence of disorder. Besides this technical
accomplishment, the results obtained with our method reveal that the distribution
of both the FTLE and the time-averaged order parameter contains useful information about the
dynamics in the stationary state and about how finite-size effects come into
play in determining the nature of stationary-state fluctuations.
In particular, we show that the distributions are significantly
skewed even in the limit $N \to \infty$, thereby implying the existence
of very different typical and atypical
behavior. Our results therefore underline the
crucial role of initial conditions in dictating the dynamical behavior,
thereby cautioning against naive reliance on 
analytically-predicted typical behavior in complex dynamical systems.

The paper is organized as follows. In Sections~\ref{sec:model}
and~\ref{sec:method}, we introduce respectively the model and the method of
study used in our paper. Numerical findings and a quantification of the
efficiency of our method are discussed in Section~\ref{sec:results}, while
conclusions are drawn in Section~\ref{sec:conclusions}. Some of the
technical details are presented in the three appendices.

\section{Model}
\label{sec:model}

Our model comprises $N$ globally-coupled oscillators \cite{note-rotor}.
The phase $\theta_i \in
[0,2\pi)$ and the angular velocity $v_i$ of the $i$-th oscillator, $i=1,2,\ldots,N$, evolve in time according to
\bea
&&\frac{{\rm d}\theta_i}{{\rm d}t}=v_i, \nonumber \\
\label{eq:eom} \\ 
&&m\frac{{\rm d}v_i}{{\rm
d}t}=\gamma(\omega_i-v_i)-\frac{K}{N}\sum_{j=1}^N \frac{\partial
u(\theta_j-\theta_i)}{\partial \theta_i}, \nonumber 
\eea
where $m$ is the moment of inertia and $\omega_i$ is the natural
frequency of the $i$-th oscillator. The $\omega_i$'s are quenched disordered
random variables obtained from a
common distribution $g(\omega)$. Here, the
second term on the right hand side of the second equation
describes the torque due to an all-to-all coupling between the oscillators, with $u(\theta)$ being the mean-field interaction potential. 
The coupling constant
$K$ in Eq. (\ref{eq:eom}) is scaled down by $N$ to have a well-defined
behavior of the associated term in the thermodynamic limit $N \to
\infty$, while the parameter $\gamma$ describes the tendency of each
individual oscillator to adapt its angular velocity to its natural
frequency. 

Mean-field interactions, such as in Eq. (\ref{eq:eom}), are a special
case (i.e., $\alpha=0$) of long-range interactions that decay asymptotically with the interparticle distance $r$ as $r^{-\alpha}$, with $0 \le \alpha \le d$ in $d$ spatial dimensions \cite{
Campa:2009,Bouchet:2010,Campa:2014,Gupta:2017}. Long-range interacting (LRI)
systems abound in nature, e.g., self-gravitating systems,
non-neutral plasma, dipolar ferroelectrics and ferromagnets,
two-dimensional geophysical vortices, wave-particle interacting systems
such as free-electron lasers, etc \cite{Campa:2014}.

In the dynamics (\ref{eq:eom}), $u(\theta)$ being even in $\theta$
\cite{note-torque}, on using $u(\theta)=u(\theta+2\pi)$, and taking $u(0)=0$ without loss of generality, a Fourier expansion yields 
\be
u(\theta)=\sum_{s=1}^\infty \widetilde{u}_s \Big[1-\cos(s\theta)\Big].
\ee
Different choices of the set $\{\widetilde{u}_s\}$ allow to describe a broad class of oscillator systems.
Here, taking $\widetilde{u}_1=1,\widetilde{u}_{s >1}=0, K>0$, we consider two 
interesting and very different limits of the dynamics:

\noindent (i)  $\gamma \to 0$: Eq. (\ref{eq:eom}) then describes the
Hamiltonian mean-field (HMF) model, a paradigmatic model of long-range
interactions \cite{Ruffo:1995,Dauxois:2002}. The dynamics is given by
\bea
\frac{{\rm d}\theta_i}{{\rm d}t}=v_i,~~
m\frac{{\rm d}v_i}{{\rm
d}t}=\frac{K}{N}\sum_{j=1}^N \sin(\theta_j-\theta_i),
\label{eq:eom-hmf}
\eea
which conserves the total energy and the total momentum $m\sum_{i=1}^N
v_i$, and admits a Boltzmann-Gibbs equilibrium stationary state. 

\noindent (ii) $m \to 0$: Eq.~(\ref{eq:eom}) then defines the Kuramoto model, a prototypical model of spontaneous synchronization in interacting
dynamical systems
\cite{Kuramoto:1975,Kuramoto:1984,Strogatz:2000,Pikovsky:2001,Acebron:2005,Strogatz:2003,Gupta:2014}:
\be
\frac{{\rm d}\theta_i}{{\rm d}t}=\omega_i+\frac{\widetilde{K}}{N}\sum_{j=1}^N
\sin(\theta_j-\theta_i);~\widetilde{K}=\frac{K}{\gamma}.
\label{eq:Kuramoto}
\ee
Note that the dynamics (\ref{eq:Kuramoto}) is first order, is
intrinsically dissipative, with the set of $\omega_i$'s continuously pumping energy into the
system. In the presence of thermal noise, the dynamics relaxes at long times to a non-equilibrium stationary state 
\cite{Gupta:2014}. 

The collective dynamics of the oscillators is quantified by the quantities 
\be
\Big(R^{(x)}_s,R^{(y)}_s\Big) \equiv \frac{1}{N}\sum_{j=1}^N
\Big(\cos(s\theta_j),\sin(s\theta_j)\Big),
\label{eq:R-defn}
\ee
in terms of which Eq.
(\ref{eq:eom}) reads
\bea
&&\frac{{\rm d}\theta_i}{{\rm d}t}=v_i,\nonumber \\
\label{eq:eom-1} \\
&&m\frac{{\rm d}v_i}{{\rm d}t}=\gamma(\omega_i-v_i)\nonumber \\
&&+K\sum_{s=1}^\infty s\widetilde{u}_s\Big[-R^{(x)}_s
\sin(s\theta_i)+R^{(y)}_s \cos(s\theta_i)\Big], \nonumber 
\eea
which shows that the dynamics is that of a single oscillator in a
self-consistent mean field due to its interaction with other oscillators. Associated with the $s$-th Fourier mode of the interaction
potential is the magnitude of the mean field given by
$R_s=\sqrt{[R^{(x)}_s]^2+[R^{(y)}_s]^2}$. Here, $R_1$ measures
phase coherence among all the oscillators, and its stationary-state
average $\langle R_1\rangle$ has been invoked as the
clustering or the magnetization order parameter in the HMF
model \cite{Ruffo:1995}, and as the 
synchronization order parameter in the Kuramoto model
\cite{Kuramoto:1975}. 

In its equilibrium state, the HMF model in the thermodynamic limit $N
\to \infty$ exhibits either a clustered (magnetized) ($\langle R_1 \rangle
> 0$) phase or
a homogeneous (unmagnetized) ($\langle R_1\rangle=0$) phase, depending on whether the energy
per oscillator $\epsilon$ is below or above a critical value $\epsilon_c \equiv 3K/4$,
respectively, with a continuous phase transition at $\epsilon_c$
\cite{Ruffo:1995}. It is convenient to depict the phases of the
oscillators as points that are moving on a unit circle under the
dynamics (\ref{eq:eom-hmf}). 
Then, in the clustered phase, the points depicting the oscillator phases are clustered together
on the unit circle at any instant of time. By contrast, in the homogeneous phase, the
points are independently and uniformly distributed on the unit circle.
In the thermodynamic limit, the
evolution of each oscillator phase becomes
equivalent to that of a pendulum in a constant gravitational
field, so that no chaos but periodic orbits are expected.
For finite $N$, however, the oscillators have the complex nonlinear dynamics
(\ref{eq:eom-hmf}) that may be characterized by a spectrum of Lyapunov exponents (LE) ${\lambda_ i}$, with $i=1,\ldots,2N$
\cite{Pikovsky:2016}. Numerical simulations and analytical
arguments for equilibrium initial
conditions have demonstrated that the maximal LE (MLE) vanishes with increasing
$N$ in the homogeneous phase. On the other hand, the MLE has a
finite value for energies just below $\epsilon_c$,
while far below $\epsilon_c$, the MLE vanishes for large $N$
\cite{Latora:1998,Firpo:1998,Latora:1999,Manos:2011,Filho:2017}. A recent study has shown that in the clustered
phase, the MLE is strictly positive for large $N$, converging to its
asymptotic value with $1/\ln N$ corrections \cite{Ginelli:2011}. 

In contrast to the HMF model, fewer studies of the chaotic properties of the
Kuramoto model exist. In the thermodynamic
limit, the model shows a stationary-state transition from a low-$\widetilde{K}$
unsynchronized phase $(\langle R_1\rangle=0)$ to a high-$\widetilde{K}$
synchronized phase $(\langle R_1\rangle > 0)$ at the critical coupling
$\widetilde{K}_c=2/[\pi g(\langle \omega \rangle)]$, where $\langle
\omega \rangle\equiv \int {\rm d}\omega~\omega g(\omega)$ is the average
frequency \cite{Kuramoto:1975}; moreover, $g''(\langle \omega
\rangle)> 0$ (respectively, $g''(\langle \omega
\rangle)< 0$) leads to a first-order (respectively, a continuous)
transition. Viewing the Kuramoto dynamics in a co-rotating
frame moving uniformly with angular frequency $\langle \omega \rangle$,
time snapshots of the oscillator phases on the unit circle in the
unsynchronized and the synchronized phase are identical to those in the
homogeneous and the clustered phase of the HMF model, respectively. In the thermodynamic limit, the Lyapunov spectrum has been analytically shown to be flat
and equal to zero below the transition, while above the transition, both
a flat and a negative branch coexist \cite{Radons:2005}. Simulations for
finite $N$ have shown the existence of a large MLE $\lambda_1(N)$ above
the transition \cite{Popovych:2005}, with $\lambda_1(N) \sim N^{-\alpha};~\alpha \le 1$
\cite{Popovych:2005,Pikovsky:2016}. 

In the rest of the paper, we use interchangeably the terms ``magnetized"
and ``synchronized" to describe the clustered phase in both the HMF model and the Kuramoto model, and the terms ``unmagnetized" and ``homogeneous" to describe
the unsynchronized phase in the two models.

Let us note that all the aforementioned studies of both the HMF model and the Kuramoto model have considered the LE's
(and mostly, the MLE) when computed over a fixed time and {\em averaged over a finite set of initial
conditions}, while
it is evidently of interest to analyze the higher moments of the MLE. Here, we are motivated to study the
chaotic properties of these systems by going beyond the average behavior of the MLE and examining in close detail the distribution
of the finite-time LE (the FTLE) with respect to different initial conditions (distributed with
the equilibrium measure in the HMF model, and with the stationary-state measure in the Kuramoto model). Our goal is to obtain the scaling with
$N$ of these distributions, with a particular emphasis on the tails 
(atypical/rare chaotic properties). Additionally, under the same
conditions and with the same focus, we obtain the
distribution of the time-averaged synchronization order parameter
(TASOP). In Appendix \ref{sec:ftle-tisop-general-definitions}, we define both the quantities of interest,
namely, the FTLE and the TASOP. In light of the highly non-linear
and many-body nature of both the HMF dynamics and the Kuramoto dynamics, the program of obtaining the desired distributions is plagued
with the usual computational challenges associated with simulations of high-dimensional non-linear systems with complex phase spaces. 
In the next section, we report on our implementation of a dedicated
importance sampling algorithm that addresses these challenges, and is particularly suited to sample rare events
in many-body interacting systems, such as the ones described by Eq.~(\ref{eq:eom}).

\section{Method}
\label{sec:method}

The main challenge in estimating the tails of the distribution of any
observable $O_\ttime$ (e.g., the FTLE, $O_\ttime\equiv
\lambda_\ttime$, and the TASOP, $O_\ttime\equiv R_\ttime$,
see Appendix \ref{sec:ftle-tisop-general-definitions})
is the rarity (exponentially small probability of occurrence) of events
that contribute to the tails. We address this challenge by employing a
Monte-Carlo Metropolis-Hastings algorithm developed in Ref.
\cite{Leitao:2017}. In essence, this algorithm uses importance sampling to draw states that are more likely to be on
the tails of the distribution of the observable, while remaining unbiased
on the averages it computes. This algorithm was developed in Ref.
\cite{Leitao:2017} to exclusively 
sample continuous phase spaces of chaotic systems. However, its
usefulness was tested only in {\em low-dimensional} (i.e., $\sim 32$
degrees of freedom) chaotic systems
evolving in {\em discrete times}. The main result of this work is to
report on a suitable adaptation and implementation of the algorithm for higher-dimensional chaotic
systems evolving in continuous times, which allows to reliably estimate
{\em even for very large $N \sim 2^{10}$} the mean, and, specifically,
the higher moments of any observable $O_\ttime$.

Let $\ver(t)$
denote a state of the system at time $t$, and $\{\ver(\tau);~\tau \in
[0,\ttime]\}$ the trajectory obtained by evolving $\ver(t)$ for time
$\ttime$ according to ${\rm d}\ver(t)/{\rm d}t={\bf F}(\ver(t))$, so
that at the end of the evolution, our observable of interest has the
value $O_\ttime(\ver(t))$. Here, $\ver(t)$ is the
$2N$-dimensional vector of phases and angular velocities of the $N$
oscillators: $r_{1,2,\ldots,N}=\theta_{i=1,2,\ldots,N}$ and
$r_{N+1,N+2,\ldots,2N}=v_{i=1,2,\ldots,N}$ in the HMF model, and is the
$N$-dimensional vector of phases of the $N$
oscillators: $r_{1,2,\ldots,N}=\theta_{i=1,2,\ldots,N}$ in the Kuramoto model.
Our objective is to
estimate the distribution $\rho(O_\ttime)$ of $O_\ttime$ for a given time
$\ttime$, a given system size $N$, and a given distribution of the
initial conditions $\ver(t)$ over a phase space region of size $\Omega$.
A knowledge of $\rho(O_\ttime)$ obviously allows to obtain
any moment of $O_\ttime$. We are particularly interested in estimating the
moments of $O_\ttime$ higher than the mean, which characterize
the tails of $\rho(O_\ttime)$.

The Metropolis-Hastings algorithm draws states
$\ver$ according to a sampling distribution $\pi(\ver)$ that favors
states yielding the tails of $\rho(O_\ttime)$ \cite{Newman:2002}. We use the so-called canonical-ensemble version of the algorithm, which has
\be
\pi(\ver)\propto e^{-\beta O_\ttime(\ver)},
\label{eq:sampling-distribution}
\ee
where $\beta$ is a parameter, a ``fictitious" temperature, that we may tune to favor
states with higher $O_\ttime$ (lower $\beta<0$) or lower $O_\ttime$ (higher
$\beta>0$). Note that with $\beta = 0$, one has a sampling distribution that is
uniform over $\Omega$, so that every initial state is equally likely to be
sampled. Hence, as far as computing the distribution of $O_\ttime$ is
concerned, a uniform sampling will merely favor the most likely values
of $O_\ttime$, that is, values of $O_\ttime$ in and around the peak of $\rho(O_\ttime)$.

In our implementation of the algorithm, we also employ the choice
(\ref{eq:sampling-distribution}). The outcome of our algorithm is a
sequence of states, $\ver \to \ver' \to \ver''\to \ldots$, that are
empirically distributed according to $\pi(\ver)$, which we use to
estimate $\rho(O_\ttime)$. Generally, the sequence is correlated, and
therefore, we evaluate the efficiency of the algorithm by estimating how
the integrated autocorrelation increases with $N$.
To obtain the states $\ver$, the algorithm employs a traditional Markov
chain with a proposal and an acceptance rate of new states.
Our implementation is based on the method developed in Ref. \cite{Leitao:2017},
with two important modifications we introduce to apply it to systems of oscillators. The first modification is the use of 
continuous time (the algorithm in Ref. \cite{Leitao:2017} was formulated
for discrete-time dynamics).  
The second modification is the way we propose the
states. Below, we systematically describe the essential steps of the
algorithm (details in Appendix \ref{app:HMF-initial-state}).

In the following, $U[a,b]$ denotes a random number uniformly distributed
in the interval $[a,b]$, and $\mathcal{N}(0,1)$ denotes a Gaussian-distributed
random number with mean zero and variance unity. Also, unless stated otherwise, the index $i$ runs over $1,2,\ldots,N_{\rm tot}$, where
$N_{\rm tot}$ is the total number of degrees of freedom: $N_{\rm
tot}=2N$ for the HMF model, and $N_{\rm tot}=N$ for the Kuramoto model.

\subsection{Preparing the initial state $\ver$}
\label{sec:prepare-initial-state}

The initial state $\ver$ of the sequence $\ver \to \ver' \to \ver''\to
\ldots$ may be prepared according to the purpose of study. For example,
for the results discussed in the paper, we sample the state according to
the equilibrium distribution in the HMF model and according to the
stationary-state distribution in the Kuramoto model. We consider for the
HMF model a net zero momentum and an
energy density above the critical value $\epsilon_c$, so that the system
is in an unsynchronized equilibrium state (We choose $K=1$ so that $\epsilon_c=3/4$). Correspondingly, the phases are uniformly and
independently distributed in $[0,2\pi)$, while the angular velocities
are Gaussian distributed; the mean of the Gaussian distribution is zero,
while the variance has to be chosen such that the prepared state has the
desired energy at which one wants to perform the study. The details of the
latter step are given in Appendix \ref{app:HMF-initial-state}. For the
Kuramoto model, we consider a Gaussian distribution with zero mean and
unit variance for $g(\omega)$; knowing the corresponding
$\widetilde{K}_c =2\sqrt{2/\pi}\approx 1.6$, we
work with $\widetilde{K}<\widetilde{K}_c$, such
that the system is in an unsynchronized stationary state;
correspondingly, the phases are uniformly and independently distributed
in $[0,2\pi)$. Note that the mentioned choices for either the HMF model
or the Kuramoto model are for illustrative purposes, and any other choice
is equally good. Once prepared, one estimates the observable $O_\ttime(\ver)$
corresponding to the initial state $\ver$. 

\subsection{Given state $\ver$, proposing a new state $\ver'$}
\label{sec:propose-new-state}

Given the state $\ver$, we propose a new state $\ver'$ by perturbing
$\ver$
as
$\ver'=\ver+\boldsymbol{\delta}$, where $\delta_i$'s are random
variables sampled independently from a half-Gaussian:
${\rm
Prob}(\delta_i)=\sqrt{2}/(A\sqrt{\pi})\exp\Big(-\delta_i^{2}/(2A^{2})\Big)$,
with $A\equiv\sigma(\ver)\sqrt{\pi/2}$.  Here, the
quantity $\sigma(\ver)$ is determined from the estimated value of $O_\ttime({\bf
r})$, as 
\begin{equation}
\sigma(\ver)\equiv\sigma_{0}\exp[-|O_\ttime({\bf r})|t^{\star}(\ver)],
\end{equation}
where $\sigma_0 \sim O(1)$ is a given parameter, and $t^{\star}$ is
related to the time needed for $O_\ttime(\ver')$ to diverge from
$O_\ttime(\ver)$.
Following Ref. \cite{Leitao:2017}, we use
\begin{equation}
t^{\star}({\bf r})\equiv{\rm
max}\Big(0,\ttime-\Big|\frac{a-1}{\beta\Big(O_\ttime^{{\rm
mp}}-O_\ttime(\ver)\Big)}\Big|\Big),
\end{equation} 
where  $O_\ttime^{{\rm mp}}$ is the most probable value of the
observable \cite{note-O-mp}, and $0<a<1$ is
a constant related to the desired acceptance rate (e.g., $a=0.01$). In this way of preparing the proposed state, it is ensured
that when $\ver$ is on the tail of the distribution $\rho(O_\ttime)$, the proposed state
is more likely to also be on the tail. On the other hand, when $\ver$ is close to the peak of
the distribution $\rho(O_\ttime)$, the proposed state would correspond to a
region around either the peak or the tail of $\rho(O_\ttime)$ \cite{Leitao:2017}.
In the case of the HMF model, whose dynamics conserves the
total energy and the total momentum, one has to implement additional
steps to make sure that the proposed state has the same momentum
(namely, zero momentum) and the
same energy as the initial state. The former is ensured by computing the
average velocity $v_{\rm avg}$ of the proposed state, as $v_{\rm avg}=\sum_{i=1}^N v_i/N$, and
then shifting the velocity of each oscillator, as $v_i\to v_i-v_{\rm avg}$;
Energy conservation is ensured by following the steps 2-6 of Appendix
\ref{app:HMF-initial-state}.

\subsection{Accepting the proposed state $\ver'$}
\label{sec:accept-new-state}

Once prepared, the proposed state $\ver'$ is accepted with a probability that ensures that the 
distribution of $O_\ttime$ corresponding to the sequence of states $\ver \to
\ver' \to \ver''\to
\ldots$ that is eventually sampled is given by Eq.
(\ref{eq:sampling-distribution}); this is achieved by ensuring that
generation and acceptance of proposed states satisfy the condition of
detailed balance~\cite{Newman:2002}: 
\bea
\pi(\ver)W(\ver\to \ver') & =\pi(\ver')W(\ver'\to \ver),
\label{eq:detailed-balance}
\eea
 where we have the transition probability
\bea
W(\ver\to\ver') & \equiv g(\ver'|\ver)P_{{\rm accept}}(\ver').
\label{eq:W-defn}
\eea
 Here, $g(\ver'|\ver)$ gives the probability of proposing the state
 $\ver'$, given state $\ver$, while $P_{{\rm accept}}(\ver')$
is the probability of accepting the state $\ver'$. Using Eq.
(\ref{eq:W-defn}) in Eq. (\ref{eq:detailed-balance}) gives 
\begin{widetext}
\bea
&&\frac{P_{{\rm accept}}(\ver')}{P_{{\rm
accept}}(\ver)}=\frac{\pi(\ver')g(\ver|\ver')}{\pi(\ver)g(\ver'|\ver)}=\exp\Big[-\beta\Big(O_\ttime({\bf
 r}')-O_\ttime({\bf
 r})\Big)\Big]\frac{\Big(1/\sigma(\ver')\Big)\exp\Big(-|{\bf
 r}-\ver'|^{2}/[\pi\sigma^{2}(\ver')]\Big)}{\Big(1/\sigma({\bf
 r})\Big)\exp\Big(-|\ver-\ver'|^{2}/[\pi\sigma^{2}({\bf
 r})]\Big)}.
 \label{eq:acceptance-probability-detail}
\eea
\end{widetext}
Corresponding to the condition
(\ref{eq:acceptance-probability-detail}), one may compute the ratio
$R\equiv\sigma(\ver)/\sigma(\ver'),$ and then the quantity
\be
r 
\equiv\log(R)-\frac{|\ver-\ver'|^{2}}{\pi\sigma^{2}(\ver)}\Big(R^{2}-1\Big)-\beta(O_\ttime({\bf
r}')-O_\ttime(\ver))\ttime.
\label{eq:acceptance-probability-definition}
\ee
Then, provided a uniformly generated random number $u\sim U[0,1]$ satisfies $\log(u)<r,$ we accept the
proposed state: $\ver\to \ver'$, record the new value of
$O_\ttime$ given by $O_\ttime(\ver'),$
and also do the updates: $t^{\star}(O_\ttime(\ver))\to
t^{\star}(O_\ttime(\ver'))$
and $\sigma(\ver)\to\sigma(\ver')$. Otherwise, we do not perform any
updates, and record the old value of $O_\ttime$ given
by $O_\ttime(\ver)$. 

Repeating the steps detailed in Sections \ref{sec:propose-new-state} and
\ref{sec:accept-new-state}, we finally generate a sequence of $M$
states ($M$ of the order of several thousands), and, correspondingly, a sequence of $M$ values of $O_\ttime$, which are
then used to construct the desired distribution $\rho(O_\ttime)$ by suitably
unbiasing \cite{Landau:2014} for the bias introduced by the choice of
(\ref{eq:sampling-distribution}) in the sampling of the states.

\section{Results}
\label{sec:results}

\begin{figure}[!ht]
\centering
\includegraphics[width=70mm]{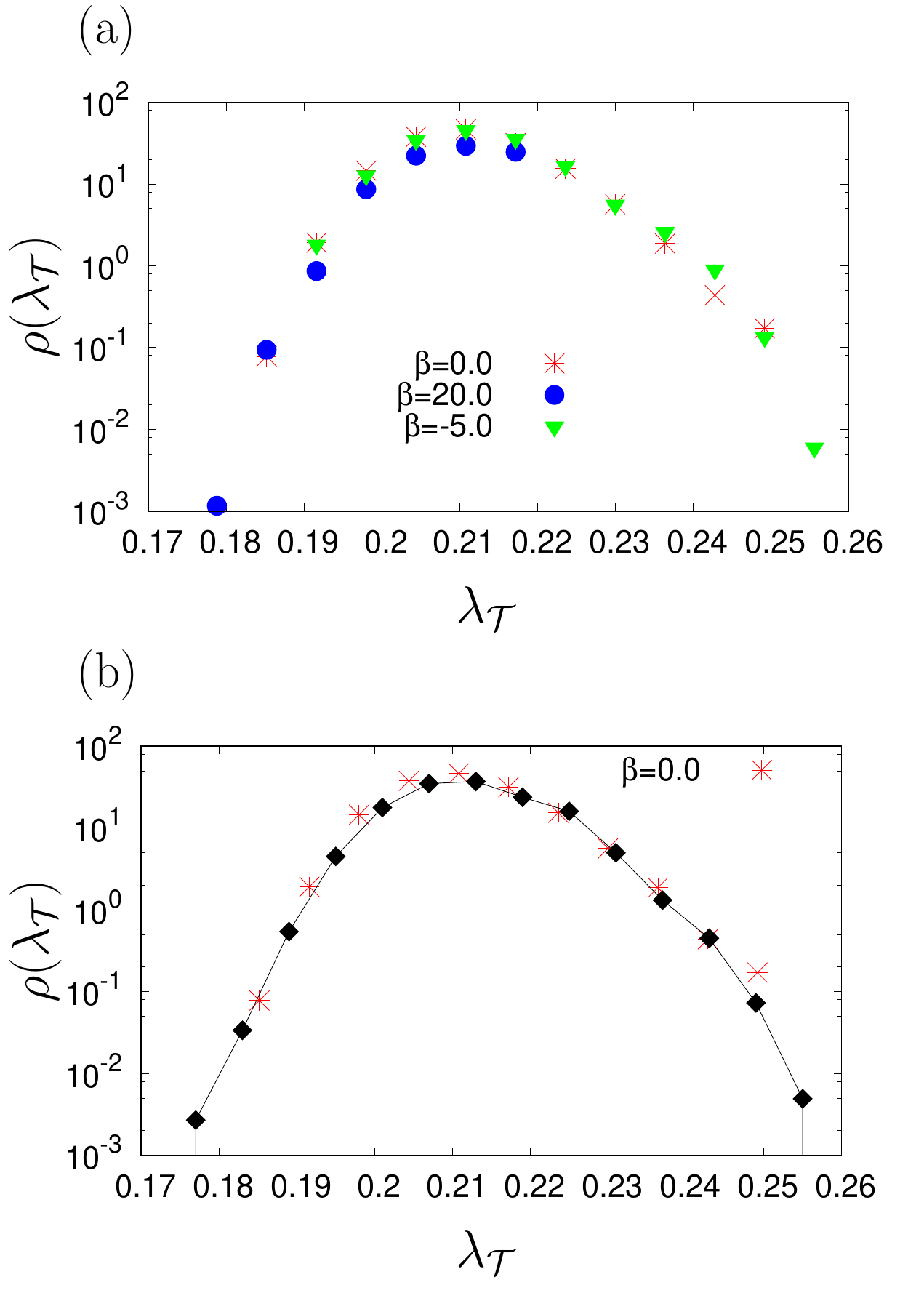}
\caption{(Color online) Our proposed Monte Carlo method leads to
improved estimation of atypical states in coupled oscillators. The
distribution of the FTLE $\lambda_\ttime$, Eq.~(\ref{eq:ftle}), was estimated numerically using
different values of the bias parameter $\beta$
in the sampling distribution~(\ref{eq:sampling-distribution}). Results
for three values of $\beta$, namely,
$\beta=-5.0,\beta=0.0,\beta=20.0$, are shown in panel (a), while a combination of the results
for these three values of $\beta$ gives the solid black (diamond) curve
in panel (b). The
system studied is the HMF model~(\ref{eq:eom-hmf}) with coupling constant $K=1$ and
energy density $\epsilon=2.0$, and we used $\ttime=10.0$.}
\label{fig:hmf-lambda-illustration}
\end{figure}

Implementing the algorithm developed in Section~\ref{sec:method}, we now
present numerical results for the two systems discussed in
Section~\ref{sec:model} -- the HMF model, Eq.~(\ref{eq:eom-hmf}), and
the Kuramoto model, Eq.~(\ref{eq:Kuramoto})
-- and for the two observables introduced in Section
\ref{sec:model} and Appendix \ref{sec:ftle-tisop-general-definitions} -- the FTLE $\lambda_\ttime$
and the TASOP $R_\ttime$.
Our first objective is to show the suitability of our algorithm in
achieving our set-out goals. Figure~\ref{fig:hmf-lambda-illustration} shows the effect of changing the bias 
$\beta$ in the sampling distribution~(\ref{eq:sampling-distribution}) on
the estimation of $\rho(\lambda_\ttime)$. On varying $\beta$, as
desired, we obtain improved estimations of $\rho(\lambda_\ttime)$ at
different range of values of $\lambda_\ttime$ (panel (a)). Combining the results
obtained with different $\beta$'s leads to an improved estimate of the
tails of $\rho(\lambda_\ttime)$, when compared to the case of $\beta=0$
(panel (b)). This success prompts us
to apply our method to investigate the $N$-dependence of the distribution.

\begin{figure*}
\includegraphics[width=160mm]{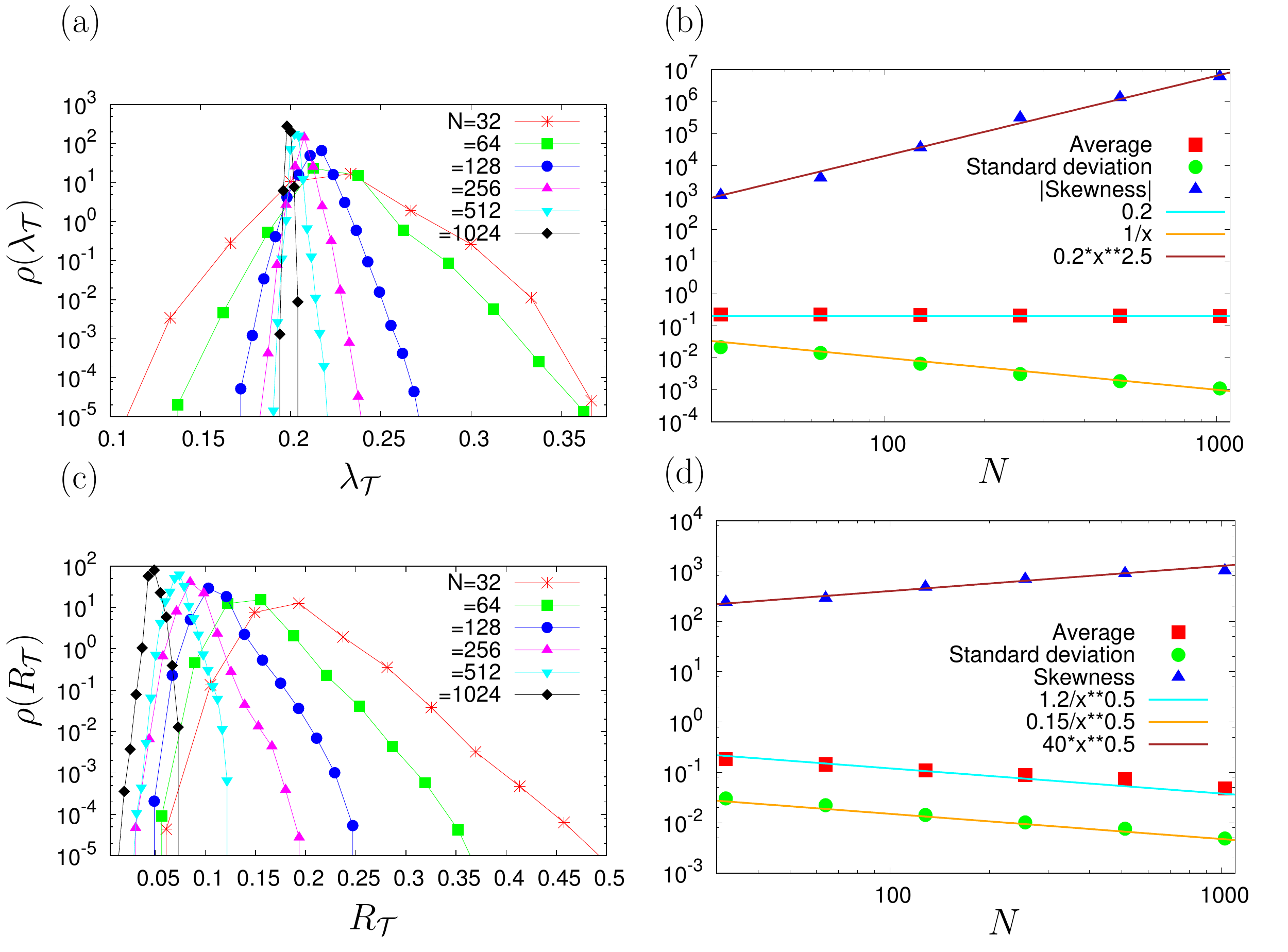} 
\caption{(Color online) Non-trivial scaling with $N$ of two observables
in the HMF model. (a) Distribution of the FTLE $\lambda_\ttime$ (with ${\cal
T}=10.0$) and (b) variation of the mean, the standard
deviation, and the skewness with the number $N$ of oscillators. (c)
Distribution of the TASOP $R_\ttime$ (with ${\cal
T}=10.0$) and (d)
variation of the mean, the standard deviation, and the skewness with the
number of oscillators. The results correspond to the HMF model~(\ref{eq:eom-hmf}) with coupling constant $K=1$ and energy density
$\epsilon=2.0$, and are obtained by performing independent
simulations with several different values of $\beta$ chosen suitably for
each $N$, and combining the results so obtained;
in each case, several thousand data points have been collected. For the
FTLE $\lambda_\ttime$, we propose and accept states using
$O_\ttime=\lambda_\ttime$ in the algorithm given in Sections 
 \ref{sec:propose-new-state} and \ref{sec:accept-new-state}. For the
 TASOP $R_\ttime$, however, we propose states using 
$O_\ttime=\lambda_\ttime$ in the algorithm given in Section
\ref{sec:propose-new-state}, and accept states using $O_\ttime=R_\ttime$
in Section \ref{sec:accept-new-state}.}
\label{fig:hmf}
\end{figure*}

\begin{figure*}
\includegraphics[width=160mm]{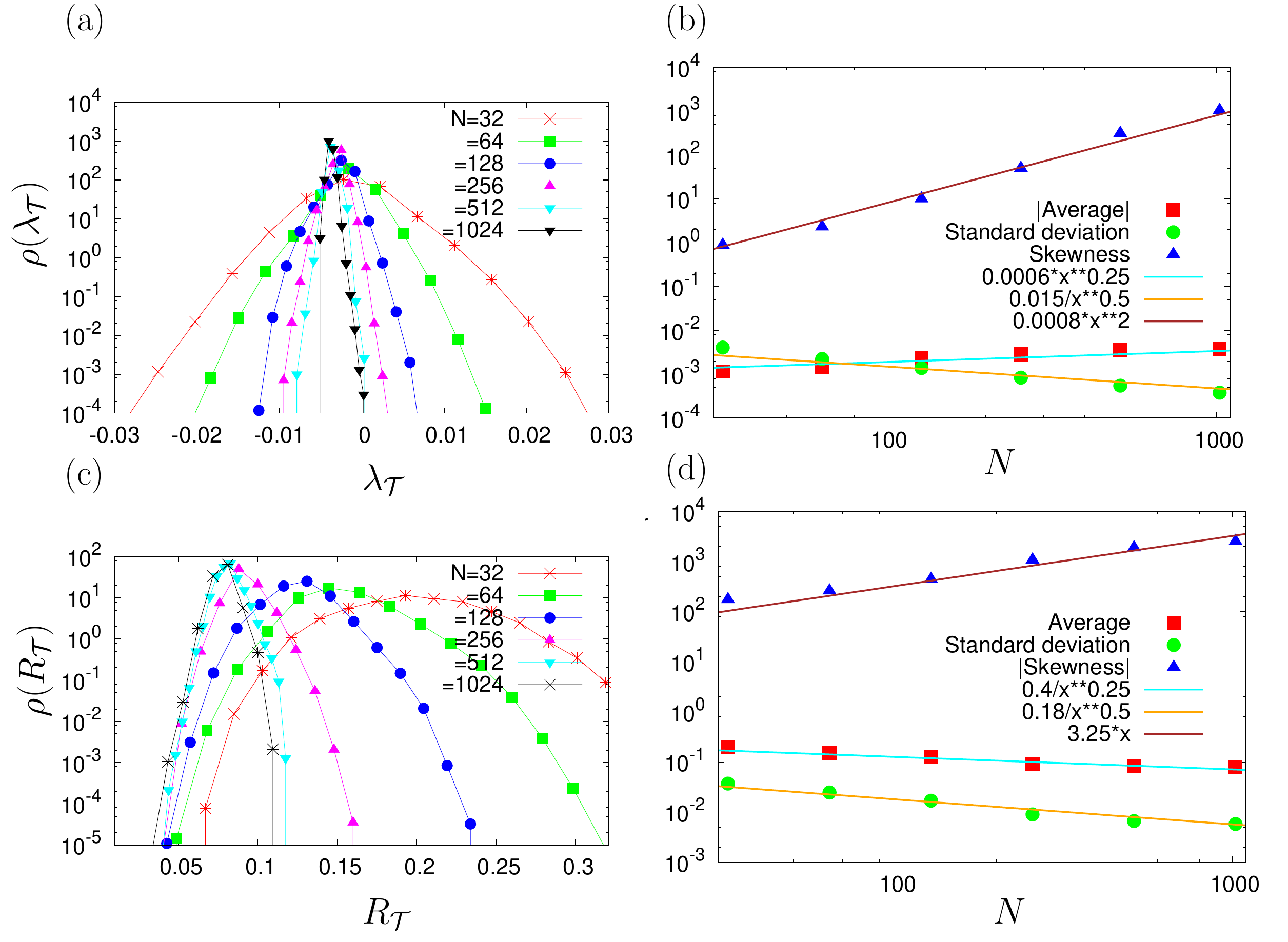}
\caption{(Color online) 
Non-trivial scaling with $N$ of two observables in the Kuramoto model.
(a) Distribution of the FTLE $\lambda_\ttime$ (with $\ttime=10.0$) and (b) variation of the mean, the standard deviation, and the skewness
with the number of oscillators $N$. (c) Distribution of the TASOP
$R_\ttime$ (with $\ttime=10.0$) and (d) variation of the mean, the
standard deviation, and the skewness with the number $N$ of oscillators.
The results correspond to the Kuramoto model~(\ref{eq:Kuramoto}) with the natural frequencies taken to be Gaussian distributed with
zero mean and unit variance, yielding
$\widetilde{K}_c=2\sqrt{2/\pi}\approx 1.6$;
here, we have $\widetilde{K}=0.25$. The results are obtained by performing independent
simulations with several different values of $\beta$ chosen suitably for each
$N$, and then combining the results so obtained;
in each case, several thousand data points have been collected. For the
FTLE $\lambda_\ttime$, we propose and accept states using
$O_\ttime=\lambda_\ttime$ in the algorithm given in Sections 
 \ref{sec:propose-new-state} and \ref{sec:accept-new-state}. For the
 TASOP $R_\ttime$, however, we propose states using 
$O_\ttime=\lambda_\ttime$ in the algorithm given in Section
\ref{sec:propose-new-state}, and accept states using $O_\ttime=R_\ttime$
in Section \ref{sec:accept-new-state}.}
\label{fig:kura}
\end{figure*}
 
The results for the HMF model are shown in Fig.~\ref{fig:hmf}. The
distributions of both the observables ($\lambda_\ttime$ and $R_\ttime$) become narrower with increasing $N$, suggesting a convergence to a single value as
$N\rightarrow \infty$. To further quantify this (expected) behaviour, we
investigate the $N$-dependence of the first three standardized moments of
the distributions (i.e., the mean, the standard deviation, and the
skewness). It is essential to employ our algorithm in the
estimation of high moments because of their values
being increasingly sensitive to the tails of the distribution.

Let us now discuss in more detail the interpretation of the results
shown in Fig.~\ref{fig:hmf}. They provide valuable insights into the properties
of the equilibrium dynamics of the HMF model that are hard
to obtain analytically due to the highly non-linear and complex nature of the
dynamics. Since we are at an energy higher than the critical
value $\epsilon_c$, the system is in an unmagnetized and chaotic steady
state, implying that one has $\lambda_{\ttime \rightarrow \infty} >0$
and $\langle R_1 \rangle=0$ in the thermodynamic limit $N \to \infty$.
From panel (a), we see here that with increase of $N$, the typical or
the most probable value of $\lambda_\ttime$ tends towards a positive value, confirming
that typical regions of the HMF model phase space are characterized by chaotic
trajectories, that is, trajectories that while starting close together
diverge from one another exponentially fast.
The results for the parameter $R_\ttime$, reported in panels (c) and (d), show that the typical value of $R_\ttime$ tends towards zero with increase of $N$,
confirming that in the large-$N$ limit, a typical state
of the system in equilibrium is unmagnetized, and during the
dynamics, the chaotic trajectories pass through sequences of states that too are unmagnetized,
being different from one another in the distribution of the points depicting the
phases of the oscillators. The main new result shown in the figure is that, even for large
$N$, atypical states with values of observables different from the
typical value are observed, as reflected by the increase in the
magnitude of the skewness with
system size $N$ (the topmost plot in panels (b) and (d)). This is
observed not only for $R_\ttime$, for which $\langle R_\ttime \rangle
\rightarrow 0$ with increase of $N$ and $R_\ttime$ as a physical
quantity satisfies $R_\ttime \ge 0$ , but also for $\lambda_\ttime$,
which  means that for finite $N$, there are a significant number of
atypical states that during the dynamics exhibit a different
chaoticity and a larger magnetization than the typical states.

Figure~\ref{fig:kura} shows that the results discussed above are
observed also in the Kuramoto model. This is significant because it
shows that our observations are valid for both first- and second-order
systems and in both conservative and dissipative dynamics. Note that the parameter values for which we investigated the Kuramoto model were such that the model has an
unsynchronized and chaotic stationary state, being therefore in a steady
state similar to the equilibrium state of the HMF model discussed in
Fig.~\ref{fig:hmf}.

On the basis of the foregoing discussions, we thus arrive at an important
conclusion regarding the dynamics of coupled oscillators, be it 
conservative as in the HMF model or dissipative as in the Kuramoto model, that stationary-state fluctuations do not die down but remain significant even for
large $N$, and that one tail of the distribution is particularly prominent with
respect to the other one. While it is definitely of urgent interest to
explain analytically the different scalings with $N$ observed in the
behavior of the moments in Figs. \ref{fig:hmf} and \ref{fig:kura}, a
roadblock is the fact that for $N$ as large as $1024$, the dynamical
trajectories of the oscillators in both the HMF model and the Kuramoto model
do not evolve independently of one other, but remain strongly coupled in
time. 

\begin{figure}
\includegraphics[width=70mm]{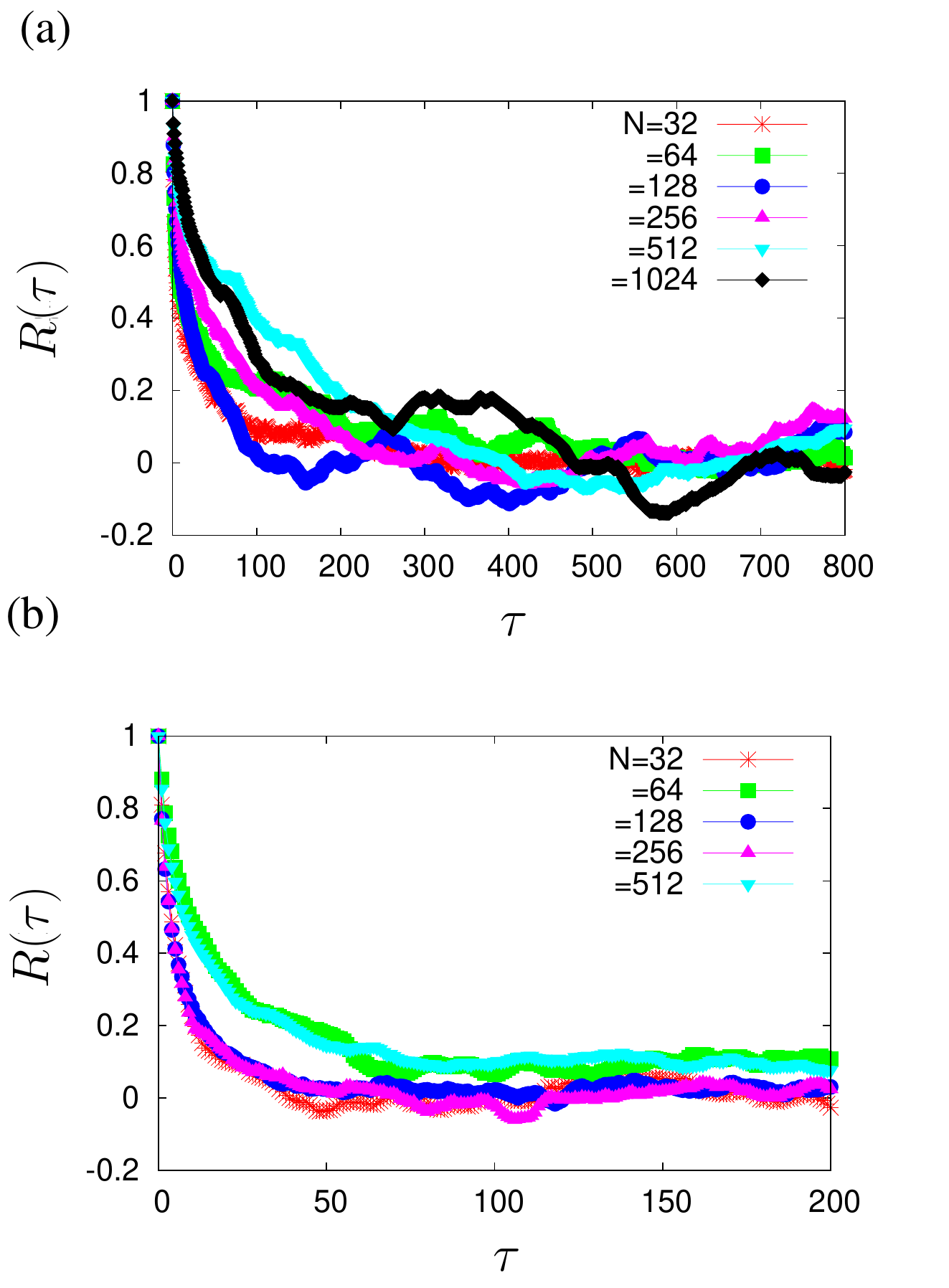}
\caption{Autocorrelation $R(\tau)$, defined in Eq.
(\ref{eq:Rtau-indicator}), for the HMF model. Panel (a) is for the
FTLE $\lambda_\ttime$, while panel (b) is for the TASOP $R_\ttime$.
The coupling constant is $K=1$, the energy density is $\epsilon=2.0$,
while we have taken $\ttime=10.0$.}  
\label{fig:Autocorrelation-indicator}
\end{figure}

Finally, we quantify the efficiency of our method. To this end, the
crucial element is the correlation of the sequence of data points
generated by our Monte Carlo algorithm~\cite{Newman:2002}, or, in other
words, the number of steps the algorithm needs to
obtain {\it independent} samples. The details of estimating this correlation
time from the generated data are described in Appendix \ref{app:efficiency}. The
results for the autocorrelation are reported in Fig.
\ref{fig:Autocorrelation-indicator}. A particularly revealing and
important observation that may be made from the figure is that the
correlation time does not (substantially) increase with increasing $N$.
This fact is further corroborated by our results on the integrated
correlation time reported in Table~\ref{tab.I}, which should be compared
with traditional (uniform sampling) method where the integrated
correlation time is expected to grow exponentially with $N$.

\begin{table}
\centering
\begin{tabular}{|c|c|}
\hline 
$N$ & $\mathscr{T}$   \\ \hline \hline
$32$ & $0.033$  \\ \hline 
$64$ & $0.055$  \\ \hline 
$128$ & $0.051$  \\ \hline 
$256$ & $0.081$  \\ \hline 
$512$ & $0.112$  \\ \hline 
1024 & $0.113$  \\ \hline 
\end{tabular}
\begin{tabular}{|c|c|}
\hline 
$N$ & $\mathscr{T}$   \\ \hline \hline
$32$ & $0.024$  \\ \hline 
$64$ & $0.050$  \\ \hline 
$128$ & $0.029$  \\ \hline 
$256$ & $0.025$  \\ \hline 
$512$ & $0.043$  \\ \hline 
$1024$ &0.19   \\ \hline 
\end{tabular}
\caption{Integrated autocorrelation $\mathscr{T}$ (see Eq.
(\ref{eq:Integrated-auto})) corresponding to Fig.
\ref{fig:Autocorrelation-indicator}, with the left (right) table
corresponding to the panel (a) (panel (b)) of Fig. \ref{fig:Autocorrelation-indicator}.
}\label{tab.I}
\end{table}

\section{Conclusions and perspectives}
\label{sec:conclusions}

In this work, we investigated the variability in the behavior of
dynamical trajectories due to different initial conditions in systems of
coupled oscillators, by developing dedicated numerical algorithms
particularly suited to study rare events in non-linear complex dynamics
of many-body dynamical systems. We focused on two representative
observables, namely, the finite-time Lyapunov exponent and a
time-averaged order parameter. With increase of the system size $N$, the distribution of such observables
becomes increasingly peaked around the value obtained in the limit $N\rightarrow \infty$,
thereby making it extremely hard to characterize for large $N$ the shape of the
distribution away from this value, with the associated computational
cost using traditional (uniform sampling) methods increasing dramatically (typically exponentially) with $N$.
Here, to show the generality of our algorithm, we studied two paradigmatic systems with
contrasting dynamics, namely, the equilibrium, second-order,
conservative dynamics of the Hamiltonian mean-field model, and the
non-equilibrium, first-order, dissipative dynamics of the Kuramoto model. We performed simulations over a wide range of system size, from $N=32$ to $N$ as large as $1024$. Our
algorithm, based on a Metropolis-Hastings Monte Carlo
Method~\cite{Newman:2002} applied earlier for low-dimensional chaotic
systems~\cite{Leitao:2017}, contains crucial adaptations necessary to
deal with coupled oscillators (e.g., extension to continuous-time
dynamics and modification of the way states are proposed).
Our numerical results confirm the accuracy
(Fig.~\ref{fig:hmf-lambda-illustration}) and efficiency
(Fig.~\ref{fig:Autocorrelation-indicator} and Table~\ref{tab.I}) of our method.

Our main numerical finding is the non-trivial convergence in the limit
$N\rightarrow \infty$ of the distribution $\rho(O_\ttime)$
to the asymptotic form. In particular, we found in all the cases of
study, namely, for the two observables in the two contrasting dynamics, that
the magnitude of the skewness of the distribution grows with $N$. This implies the relative importance
of one tail of the distribution over the other with
increase of $N$, despite reduced
fluctuations (i.e., with the standard deviation $\to 0$). These observations suggest that even for large
$N$, states yielding values of observables substantially deviated from the
typical value occur with significant probabilities. To 
understand theoretically the scaling of the different moments with $N$
observed in Figs. \ref{fig:hmf} and \ref{fig:kura} is a particularly challenging task
left for future studies.

The successful application of our method demonstrated in this work
suggests the exciting possibility to apply our algorithm to more complex
dynamics, e.g., to investigate the properties of the out-of-equilibrium
quasi-stationary states \cite{Campa:2014} in the HMF model, the subtle dependence of
the dynamics on the realizations of the natural
frequencies $\omega_i$'s in the Kuramoto model, the existence of the so-called chimera
states \cite{Panaggio:2015} in oscillator systems in which they appear, and also in determining the distribution of Lyapunov exponents in other relevant long-range systems such as the
self-gravitating ring model \cite{Tatekawa:2015}. 

\section{Acknowledgements}
SG acknowledges fruitful discussions with T. Manos, A. Politi, and S. Ruffo. JL acknowledges funding from grant SFRH/BD/90050/2012 (FCT Portugal) and fruitful discussions with J. V. Parente Lopes.

\appendix
\section{Observables: The FTLE $\lambda_\ttime$ and the TASOP $R_\ttime$}
\label{sec:ftle-tisop-general-definitions}

\subsection{The finite-time Lyapunov exponent $\lambda_\ttime$ (FTLE)}
\label{app:FTLE-defn}

We define here the FTLE $\lambda_\ttime$ corresponding to the dynamics
(\ref{eq:eom}). The definition may be suitably modified for the HMF and
the Kuramoto model by taking the appropriate limits of the dynamics discussed in Section
\ref{sec:model}.
Let $\ver(t)$ denote a state of the system at time $t$, and let
$\{\ver(\tau);~\tau \in
[0,\ttime]\}$ be the trajectory obtained by evolving a given initial
state for time $\ttime$ according to the dynamics of $\ver(t)$. Here, $\ver(t)$ is the
$2N$-dimensional vector of phases and angular velocities of the $N$
oscillators: $r_{1,2,\ldots,N}=\theta_{i=1,2,\ldots,N}$ and
$r_{N+1,N+2,\ldots,2N}=v_{i=1,2,\ldots,N}$.
Corresponding to the dynamics 
\be
\frac{{\rm d}\ver(t)}{{\rm d}t}={\bf F}({\bf
r}(t)),
\label{eq:eom-phase-space}
\ee
a set of infinitesimal
displacements $\{\delta r_i(t)\}$ from a given phase space point $\ver(t)$
at time $t$ will have its evolution described by the following linearized
equations of motion:
\be
\delta r_i(t)=\sum_{j=1}^{2N}\frac{{\rm d}{F}_i}{{\rm d}r_j}\Big|_{{\bf
r}(t)}\delta r_j(t);~~i,j=1,2,\ldots,2N,
\label{eq:eom-tangentspace}
\ee
where the elements $\frac{{\rm d}F_i}{{\rm d}r_j}\Big|_{{\bf
r}(t)}$ define the Jacobian matrix $J_{ij}$ evaluated at $\ver(t)$. For the dynamics (\ref{eq:eom}), one has
\begin{widetext}
\bea
F_i=\left\{
\begin{array}{ll}
               r_{i+N} \mbox{~for  $1\le i \le N$}, & \\
               \gamma(\omega_{i-N}-r_i)/m+(K/m)\sum_{s=1}^\infty s\widetilde{u}_s\Big[-R^{(x)}_s
\sin(sr_{i-N})+R^{(y)}_s \cos(sr_{i-N})\Big] & \nonumber \\
               \mbox{~for  $N+1\le i \le 2N$}, & \\
               \end{array}
        \right. \\ 
\eea
\bea
J_{ij}=\left\{
\begin{array}{ll}
               1 \mbox{~for  $1\le i \le N, N+1 \le j \le 2N$}, & \\
               -(K/(Nm))\sum_{s=1}^\infty s\widetilde{u}_s
               \cos(s(r_{i-N}-r_{j})) 
               \mbox{~for  $N+1\le i \le 2N, 1 \le j \le N,i\ne j$}, & \\
               (K/m)\sum_{s=1}^\infty s\widetilde{u}_s
               \Big(R^{(x)}_s\cos(sr_{j})+R^{(y)}_s\sin(sr_{j})-\frac{1}{N}\Big)
                              \mbox{~for  $N+1\le i \le 2N, 1 \le j \le
                              N,i=j$}, & \nonumber \\
               0                \mbox{~otherwise}. &
               \end{array}
        \right. \\
\eea
\end{widetext}
Starting with a given set $\{\ver(t),\delta \ver(t)\}$ at time $t$, and
evolving under Eqs. (\ref{eq:eom-phase-space}) and (\ref{eq:eom-tangentspace}) for a
total time $\ttime$, the FTLE $\lambda_\ttime$ is defined as
\be
\lambda_\ttime(\ver(t))\equiv \frac{1}{\ttime}\log
\frac{d(t+\ttime)}{d(t)},
\label{eq:ftle}
\ee
with $d(t) \equiv \sqrt{\sum_{i=1}^{2N} [\delta r_i (t)]^2}$ the metric
distance calculated from the infinitesimal displacements measured from
the instantaneous location $\ver(t)$ at time $t$.
The FTLE $\lambda_\ttime$ is thus a measure of the finite-time (i.e., in time $\ttime$) exponential divergence of
different trajectories starting close to $\ver(t)$ at time $t$. 

We used the following algorithm to efficiently compute $\lambda_\ttime$. The starting point is the generation of the perturbations
$\delta r_i$ for $i=1,2,\ldots,N_{\rm tot}$; the obtained values are then scaled as
$\delta r_i \to\delta r_i/d$, with
$d\equiv\Big(\sum_{i=1}^{N_{\rm tot}}(\delta r_i)^{2}\Big)^{1/2}$.
Starting with the given set $\{r_i\}$ and $\{\delta
r_i\}$, their values are evolved
in time \cite{Integrator} according to the equations of motion of the
model at hand for a total time $\ttime$, by computing at
every time interval $\tau$ the quantity
$d_{\alpha}\equiv\Big(\sum_{i=1}^{N_{\rm tot}}(\delta r_{i}^{(\alpha)})^{2}\Big)^{1/2},$
with $\alpha=1,2,\ldots,\ttime/\tau,$ and by rescaling the
perturbations as $\delta r_i^{(\alpha)}\to\delta
r_i^{(\alpha)}/d_{\alpha}$. Here, $N_{\rm tot}$ is the total number of degrees of freedom: $N_{\rm
tot}=2N$.
Then $\lambda_\ttime$ is estimated as 
\bea
\lambda_\ttime(\ver) &
\equiv\frac{1}{\ttime}\sum_{\alpha=1}^{\ttime/\tau}\ln d_{\alpha}.\label{eq:lambda-computation-formula}
\eea

\subsection{The time-averaged synchronization order parameter $R_\ttime$ (TASOP)}
\label{app:TASOP-defn}

A time evolution for time $\ttime$ while starting from a given phase
space point $\ver(t)$ yields the time-averaged synchronization order
parameter $R_\ttime$ (TASOP) defined as
\be
R_{s,\ttime}(\ver(t))\equiv \frac{1}{\ttime} \int_t^{t+{\cal
T}} {\rm d}t'~||\Big(R^{(x)}_s(t'),R^{(y)}_s(t')\Big) ||,
\label{eq.RT}
\ee
with $||(x,y)|| \equiv \sqrt{x^2+y^2}$, and $R^{(x),(y)}_s(t')$
corresponding to the phase space point $\ver(t')$ is defined in Eq.
(\ref{eq:R-defn}). In this work, we will specifically
consider $R_\ttime(\ver(t))\equiv R_{1,\ttime}(\ver(t))$. Note that
$R_\ttime$ is a positive quantity: $R_\ttime \ge 0$.

\section{Preparation of the initial state ${\bf
r}\equiv\{\theta_{i},v_{i}\}$
with fixed energy density $\epsilon$ and net momentum zero for the HMF model}
\label{app:HMF-initial-state}

The relevant steps are
\begin{enumerate}
\item Assign independently for $i=1,2,\ldots,N$ the $\theta_{i}$'s as
$\theta_{i}\sim U[0,2\pi]$. 
\item Compute the potential energy
$V(\{\theta_{i}\})=\frac{K}{2N}\sum_{i,j=1}^{N}[1-\cos(\theta_{i}-\theta_{j})]$,
and thence the allowed value of the kinetic energy per oscillator for the
given realization of the $\theta_{i}$'s as $T_{{\rm
allowed}}(\{\theta_{i}\})\equiv\epsilon-V(\{\theta_{i}\})/N$. 
\item Assign independently for $i=1,2,\ldots,N$ the angular velocities
$v_{i}$'s as $v_{i}\sim\sqrt{2T_{{\rm
allowed}}((\theta_{i}\})}\mathcal{N}(0,1)$.
In this way, $\sum_{i=1}^{N}v_{i}$ is very close to zero (with corrections
that decrease with increasing $N)$, but is not exactly so, and also
$\sum_{i=1}^{N}v_{i}^{2}/2$ is very close to the allowed value $NT_{{\rm allowed}}(\{\theta_{i}\})$
of the kinetic energy (with corrections that decrease with increasing
$N)$, but is not exactly so. 
\item Compute the average velocity $v_{\rm
avg}\equiv\sum_{i=1}^{N}v_{i}/N,$ and transform for $i=1,2,\ldots,N$
the $v_{i}$'s as $v_{i}\to v'_{i}=v_{i}-v_{\rm avg},$ thereby ensuring that
$\sum_{i=1}^{N}v'_{i}=0$. From now on, let us drop the prime for
notational convenience. 
\item Compute the actual kinetic energy $T_{{\rm
actual}}(\{\theta_{i},v_{i}\})$
for the set of $v_{i}$'s at hand, while we already have the value
of the potential energy $V(\{\theta_{i}\})$ from step 2. We then
compute the quantity
$\alpha\equiv\sqrt{|N\epsilon-V(\{\theta_{i}\})|/T_{{\rm
actual}}(\{\theta_{i},v_{i}\})}$. 
\item Scale for $i=1,2,\ldots,N$ the $v_{i}$'s as
$v_{i}'\equiv\alpha v_{i}$.
This ensures that
$\sum_{i=1}^{N}v_{i}^{2}/2=N\epsilon-V(\{\theta_{i}\})=NT_{{\rm
allowed}}(\{\theta_{i}\})$,
as desired.
\end{enumerate}

\section{Efficiency of the algorithm}
\label{app:efficiency}

The efficiency of our algorithm depends critically on the correlation
time of the generated sequence of the values of the observable, which
may be quantified as follows. For a given $N$, the
distribution of $O_\ttime$ obtained by uniform sampling gives an
estimate of the most probable value $O^{\rm mp}_\ttime$. We then
choose a value $O^{\rm tail}_\ttime$ of $O_\ttime$ on the tail, say, at a distance $2\sigma$
from $O^{\rm mp}_\ttime$, where $\sigma$ is the standard deviation of
the distribution of $O_\ttime$. We then employ the importance sampling
scheme detailed in the paper to obtain a time series $\{O_{\cal
T}^{(i)}\}_{1\le i\le M}$ and thence, a biased distribution centered
around $O^{\rm tail}_\ttime$. Next, knowing the full width at half
maximum of the biased distribution, we define an indicator variable $I_{i}$
that takes the value of unity provided the $i$-th value $O_\ttime^{(i)}$
lies within the full width and takes the value zero otherwise. The autocorrelation is then computed as
\be
R(\tau)
\equiv\frac{\sum_{i=1}^{M-\tau}\Big(I_{i}-\overline{I}\Big)\Big(I_{i+\tau}-\overline{I}\Big)}{(M-\tau)\sigma_{I}^{2}};~~0\le
\tau \le \frac{M}{2},\label{eq:Rtau-indicator}
\ee
where $\overline{I}\equiv\frac{1}{M}\sum_{i=1}^{M}I_{i},$ and $\sigma_{I}^{2}\equiv\frac{1}{M}\sum_{i=1}^{M}(I_{i}-\overline{I})^{2}$.
The integrated autocorrelation is defined as
\be
\mathscr{T} \equiv \frac{2}{M}\sum_{\tau=1}^{M/2}|R(\tau)|.
\label{eq:Integrated-auto}
\ee



\begin{thebibliography}{99}
\bibitem{note-rotor}In
presence of a finite moment of inertia, an oscillator is more
traditionally referred to as a rotor; here, however, we continue to refer to
it as an oscillator.
\bibitem{note-torque}This ensures that the internal torque on the $i$-th oscillator
due to the $j$-th one is balanced by that on the $j$-th oscillator due to the
$i$-th one.
\bibitem{note-O-mp}This value may be
easily estimated by performing a $\beta=0$ simulation.
\end{thebibliography}
\end{document}